\documentclass[aps,pra,showpacs,reprint,amsmath,amssymb,floatfix,groupedaddress]{revtex4-1}

\usepackage{graphicx}
\usepackage{epsfig}

\begin{document}

\title{Production of Ultracold Molecules with Chirped Nanosecond Pulses: Evidence for Coherent Effects}

\author{J. L. Carini$^{1}$}
\thanks{These authors contributed equally to this work.}
\author{J. A. Pechkis$^{1,2}$}
\thanks{These authors contributed equally to this work.}
\author{C. E. Rogers III$^{1}$}
\author{P. L. Gould$^{1}$}
\author{S. Kallush$^{3}$}
\author{R. Kosloff$^{4}$}
\affiliation{$^{1}$Department of Physics, University of Connecticut, Storrs, Connecticut 06269, USA \\ $^{2}$ Currently with the Joint Quantum Institute, University of Maryland, College Park, MD 20742, USA \\ $^{3}$Department of Physics and Optical Engineering, ORT Braude, P.O. Box 78, Karmiel, Israel \\ $^{4}$Department of Physical Chemistry and the Fritz Haber Research Center for Molecular Dynamics, The Hebrew University, 91094, Jerusalem, Israel}

\date{\today}

\begin{abstract}
We use frequency-chirped light on the nanosecond time scale to produce ultracold $^{87}$Rb$_{2}$ molecules in the lowest triplet state via the process of photoassociation.  Comparing to quantum simulations of the molecular formation, we conclude that coherent stimulated emission plays an important role and is primarily responsible for the significant difference observed between positive and negative chirps.
\end{abstract}

\pacs{32.80.Qk, 37.10.Mn, 34.50.Rk}

\maketitle


\section{Introduction}
Applying the concepts of coherent control to the manipulation of ultracold systems is a topic of considerable current interest. Coherent control \cite{Rice00, Shapiro03} usually involves internal degrees of freedom, such as molecular vibration and rotation, while cooling and trapping techniques \cite{Metcalf99} deal with external degrees of freedom. The time scales are usually quite different as well: coherent control is typically done with ultrafast lasers while motion at ultralow temperatures is very slow. A particularly noteworthy  convergence of these two fields is the formation of ultracold molecules from ultracold atoms by the process of photoassociation \cite{Jones06} (PA). This free-bound transition is a simple binary reaction starting with a narrow range of continuum energies, so coherence can be expected to play an important role \cite{Koch12}. 

In recent years, there have been many proposals for coherent control of PA with shaped ultrafast pulses \cite{Koch12}. This has been motivated in large part by the desire to form ultracold molecules \cite{Krems09} with high efficiency and state specificity for their many potential applications in quantum information, precision spectroscopy, ultracold chemistry, and quantum dipolar systems. So far, experimental progress towards coherently controlled PA has been limited to control of the photodestruction of already existing ultracold molecules \cite{Salzmann06, Brown06} and the observation of coherent transients in PA with femtosecond pulses \cite{Salzmann08, McCabe09}. In recent work, we have used frequency-chirped light on the nanosecond time scale to coherently control a closely related process, laser-induced inelastic collisions. Because our nanosecond pulses are well matched to the long-range motion of the colliding atoms, we have seen that the collision rate depends not only on the chirp direction \cite{Wright07}, but also on the shape of the chirp \cite{Pechkis11}. In the present work, we apply our chirped pulses to the process of PA and directly detect the resulting ground-state molecules. We see a dependence of the formation rate on chirp direction, in agreement with quantum simulations of the PA dynamics. These simulations reveal that despite the presence of spontaneous emission, a significant portion of the chirp dependence arises from a coherent effect: stimulated emission into a specific high vibrational level. 

\section{Experiment}
In the experiment \cite{Wright07}, we illuminate ultracold $^{87}$Rb atoms with nanosecond-scale pulses of frequency-chirped light, forming long-range excited-state $^{87}$Rb$_{2}$ molecules via PA. These excited molecules subsequently radiatively decay into high vibrational levels (v'') of the a $^{3}$$\Sigma$$_{u}^{+}$ metastable state which are detected by resonantly-enhanced multiphoton ionization (REMPI).
The ultracold atoms are provided by a magneto-optical trap (MOT) operated in the phase-stable configuration and loaded continuously by a slow atomic beam emanating from a separate source MOT. The atomic temperature and peak density are $\sim$150 $\mu$K and $\sim$5x10$^{10}$ cm$^{-3}$, respectively.

The frequency-chirped light is produced by rapidly varying the injection current of an external-cavity diode laser with a 240 MHz arbitrary waveform generator (AWG) and the chirp shape is measured by a heterodyne technique. The AWG waveform is approximately a 5 MHz triangle wave, but with programmed adjustments to produce approximately linear positive and negative chirps with equal slopes during the pulse. Each chirp covers approximately 1 GHz in 100 ns and is centered on the $^{87}$Rb$_{2}$ PA transition, a strong line located 7.79 GHz below the $5S_{1/2} (F=2) \rightarrow 5P_{3/2} (F'=3)$ asymptote and determined to have 0$_{g}^{-}$ character \cite{Kemmann04}. To minimize the amplitude modulation, the chirped light is used to injection lock a separate 150 mW slave diode laser \cite{Wright04}. A sequence of 40 ns FWHM pulses of the chirped light is generated by switching with an acousto-optical modulator (AOM). The timing selects either the positive or negative chirps.

REMPI detection of the resulting Rb$_{2}$ molecules is performed with 5 ns, 4 mJ pulses from a Nd-YAG pumped pulsed dye laser tuned to  $\lambda$=601.9 nm and focused to $\sim$3 mm diameter at the MOT location. Based on previous work \cite{Gabbanini00, Kemmann04}, this light ionizes high-v'' levels of the a $^{3}$$\Sigma$$_{u}^{+}$ state which are expected to be populated by our PA to long-range excited states. The REMPI spectrum is similar to that from molecules produced by MOT light, and is dominated by a broad feature centered at 601.9 nm. Individual high-lying v'' levels are not resolved due to the 0.2 cm$^{-1}$ pulsed laser bandwidth. The resulting Rb$_{2}^{+}$ ions are accelerated to a Channeltron ion detector and distinguished from Rb$^{+}$ ions by their time of flight.  The timing of the experiment is as follows. A sequence of up to 5x10$^{4}$ chirped (or unchirped) pulses is applied to the trapped atoms and 25 $\mu$s later, the REMPI pulse fires and the ions are detected. The entire cycle is repeated at 10 Hz. The MOT beams are extinguished during a 50 $\mu$s window centered on the REMPI pulse to avoid ionization of excited atoms by the REMPI light. We use a train of chirped PA pulses, so in order to obtain the actual molecular formation rate R, we must account for the loss of molecules during this PA time window. There is slow photodestruction of a $^{3}$$\Sigma$$_{u}^{+}$ molecules by subsequent chirped pulses at a time-averaged rate $\Gamma_{PD}$ as well as their escape from the detection region at a rate $\Gamma_{esc}$. The number N of detectable molecules varies with time according to:\begin{equation}
\setcounter{equation}{1}
N(t) = \frac{R}{\Gamma_{PD}+\Gamma_{esc}}(1 - e^{-(\Gamma_{PD}+ \Gamma_{esc})t}).	
\end{equation}
We determine $\Gamma_{esc}$ = 108(7) s$^{-1}$ by measuring the exponential decay of the REMPI signal from MOT-produced molecules (i.e., without chirped PA light) as the REMPI pulse is delayed within a fixed 7 ms window following extinction of the MOT light. We determine $\Gamma_{PD}$ and R by measuring the REMPI signal as a function of the length of the PA window (i.e., the number of chirped pulses) and fitting to Eq. 1 as shown in Fig. 1(a). For this case, $\Gamma_{PD}$ $\sim$200 s$^{-1}$ for the positive chirp, implying a photodestruction probability of $\sim$4x10$^{-5}$ per pulse. We find that $\Gamma_{PD}$ is linear in intensity and depends somewhat on chirp direction. 

\begin{figure}
    \includegraphics[width=8.3cm]{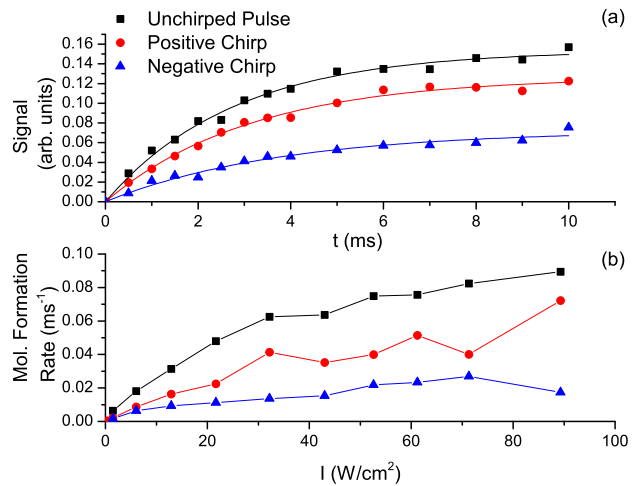} 
    \caption{(Color online) (a) Rb$_{2}^{+}$ REMPI signals vs. photoassociation time for unchirped, positively chirped, and negatively chirped pulses, along with fits to Eq. 1. The pulse repetition rate is 5x10$^{6}$ pulses/s and the peak intensity of the pulses is fixed at 32.2 W/cm$^{2}$.  (b) Rb$_{2}$ formation rate vs. intensity for the various chirps.}
\end{figure}

The quantity of interest is the formation rate R shown in Fig. 1(b). These values are derived from the fits to curves such as Fig. 1(a) for various intensities. For each chirp direction, R increases with intensity, but exhibits some degree of saturation. The important point is that there is a pronounced dependence on chirp: the positive chirp has a rate higher than the negative chirp, but lower than the unchirped case.

\section{Simulations}
To model the ultracold collisional dynamics, we solve the time-dependent Schr$\ddot{o}$dinger equation. The dressed-state Hamiltonian reads:\begin{equation}
\setcounter{equation}{2}
\hat{H}=
\begin{pmatrix}
\hat{T}+\hat{V}_{gJ} & \hbar \Omega_{0}(t) &  \hbar \Omega_{1}(t) \\ \hbar \Omega_{0}^{*}(t) & \hat{T}+\hat{V}_{0}+\Delta & 0 \\ \hbar \Omega_{1}^{*}(t) & 0 & \hat{T}+\hat{V}_{1}+\Delta 
\end{pmatrix},
\end{equation} where $\hat{T}$ is the kinetic energy operator and  $\hat{V}_{j}$ (j = g, 0 or 1 for the a $^{3}$$\Sigma$$_{u}^{+}$, $0_{g}^{-}$ and $1_{g}$ electronic states) are the internuclear potentials computed in \cite{Kosh10} with coefficients from \cite{Gutterres} adjusted for the proper scattering length \cite{Roberts, Geltman}. The two excited states correspond to the assignment of \cite{Kemmann04}.  $\Delta$ is the central detuning of the light from the asymptote [$5S_{1/2} (F=2) + 5P_{3/2} (F'=3)$]. The $0_{g}^{-}$ detuning was shifted to yield the correct experimental spacing of $\sim$0.6 GHz and to be centered on v'=78, in accordance with \cite{Kemmann04}. For partial waves other than s, we add a rotational barrier $V_{J} = J(J+1)/2mR^{2}$ to the a $^{3}$$\Sigma$$_{u}^{+}$ potential.

 The time-dependent couplings between the a $^{3}$$\Sigma$$_{u}^{+}$ (g) and excited (j=0,1) states due to the chirped pulse are given by: \begin{equation}
\setcounter{equation}{3}
\hbar \Omega_{j} = \mu_{gj} \epsilon_{0}e^{[-(\frac{t-t_{center}}{2\sigma})^{2}+i\widetilde{\omega}(t)(t-t_{center})]},
\end{equation} where $\mu_{gj}$ are the transition dipole moments (including Franck-Condon factors), $\epsilon_{0}$ is the peak electric field, $\sigma$ is the pulse width, $t_{center}$ is the center of the intensity pulse, and $\widetilde{\omega}(t)$ are instantaneous frequency offsets from $\Delta$ derived from smoothed interpolations of the heterodyne signals \cite{Carini12}.  

To enable efficient computation for nanosecond timescales, the simulations are performed within the basis of vibrational levels calculated on a time-independent mapped Fourier grid \cite{Kallush06, Kallush07}. A limited bandwidth is then taken to represent each of the vibrational Hamiltonians: $\sim$15 GHz for the $0_{g}^{-}$ and $1_{g}$ excited states; and 278 GHz (16 MHz) for the a $^{3}$$\Sigma$$_{u}^{+}$ bound (scattering) manifold.  We have verified that this representation is sufficient by extending the basis sets and checking convergence.  The initial single state is a box-normalized scattering state at E$_{0}$ = $k_{B}$T, where T = 150 $\mu$K is the sample temperature.

As described in \cite{Carini12}, spontaneous decay is taken into account by adding multiple sink channels corresponding to decay from each of the excited potentials into various a $^{3}$$\Sigma$$_{u}^{+}$ vibrational levels. Note that this model precludes the possibility of multiple incoherent excitations. However, as discussed below, almost all of the population that decays into detectable levels is far from resonance and will not participate in subsequent dynamics. 

The computation gives the probability of molecular production, $P_{E_{0},J}$, for a given initial box-normalized state and a given partial wave, J. Representing the Hamiltonian (Eq. 2) on a finite grid in coordinate space allows us to divide the trap volume, V, into many smaller boxes of volume $\nu_{box}$, whose number is larger then the number of atoms, N.  Then, following \cite{Kosh06}, the number of molecules, $N_{mol,J}$, is computed by:\begin{equation}
\setcounter{equation}{4}
N_{mol,J} = \frac{N^{2}}{2} \frac{\nu_{box}}{V} \langle \hat{P_{e,J}} \rangle _{T, box} = \frac{\pi^{2} \hbar^{3} Nn P_{E_{0},J}}{\mu^{3/2} \sqrt{E_{0}} \left. \frac{dE}{dn}\right|_{E_{0}}} \frac{}{},
\end{equation} where n is the density, $\mu$ is the reduced mass, and $\left. \frac{dE}{dn}\right|_{E_{0}}$ is the density of energy states evaluated at E$_{0}$.  To find the molecular formation rate P$_{J}$ for each peak intensity I$_{0}$, we then multiply by the number of chirps per ms and spatially average over the density distribution in the trap and intensity profile of the photoassociation laser. Following \cite{Kosh06}, we find the overall molecular formation rate by summing over all the partial waves necessary for convergence:\begin{equation}
\setcounter{equation}{5}
P(I_{0}) = \sum^{5}_{J=0} (2J+1) P_{J}(I_{0}).
\end{equation}

\begin{figure}
    \includegraphics[width=8.3cm]{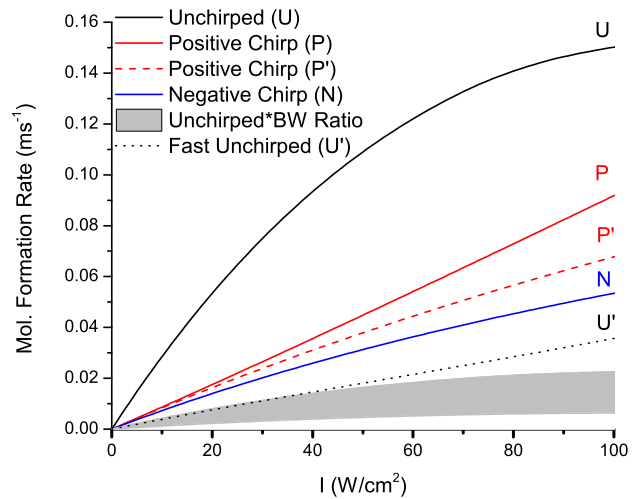} 
    \caption{(Color online) Simulated molecular formation rates vs. intensity for unchirped (U), positively chirped (P), and negatively chirped (N) pulses. The dashed curve (P') is for the positive chirp, but excluding the coherent contribution to v''=39. Also shown as the shaded region is the range of results for the unchirped pulses scaled by the ratio of the unchirped bandwidth to the chirped bandwidth (see text). The dotted curve U' is for a pulse shorter by a factor of 48.}
\end{figure}

\begin{figure}
\includegraphics[width=8.3cm]{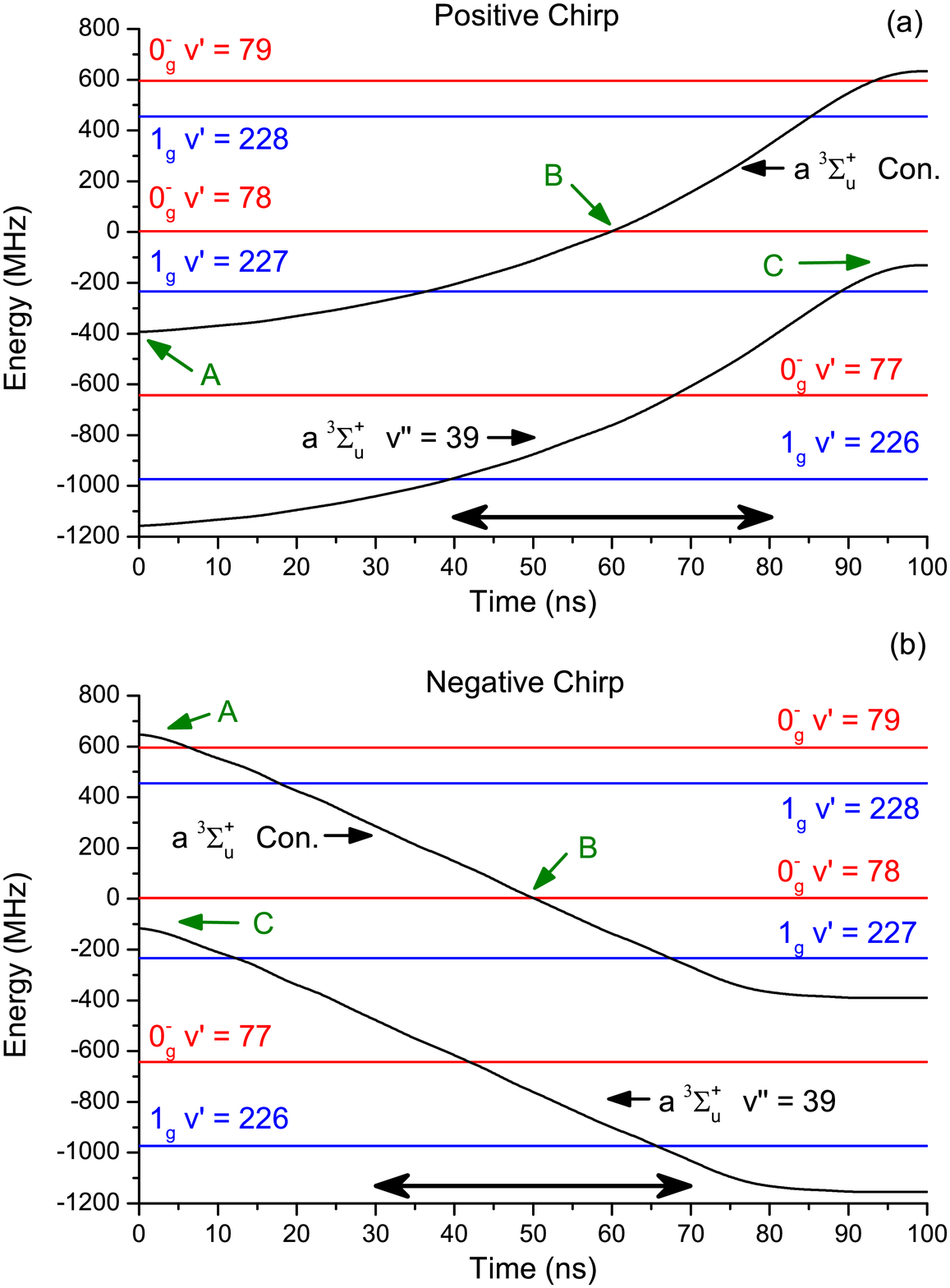}
    \caption{(Color online) Evolution of the molecular levels during the positively chirped (a), and negatively chirped (b), pulses. The horizontal lines are the relative energies of the vibrational levels of the excited 0$_{g}^{-}$ and 1$_{g}$ molecular states, while the energies of the a $^{3}$$\Sigma$$_{u}^{+}$ zero-energy continuum and the a $^{3}$$\Sigma$$_{u}^{+}$ level, with the energy of the chirped photon added, are represented by the upper and lower black curves, respectively. In this picture, a curve crossing indicates when the light is resonant with the corresponding transition. Ground-excited couplings are not included in these plots. Double-ended arrows indicate the pulse widths (FWHM). The point labeled A is the initial continuum state, B indicates resonance with the 0$_{g}^{-}$ (v'=78) level, and C indicates the approach to resonance with the transition from 0$_{g}^{-}$ (v'=78) down to a $^{3}$$\Sigma$$_{u}^{+}$ (v''=39). }
\end{figure}

\begin{figure}
    \includegraphics[width=8.3cm]{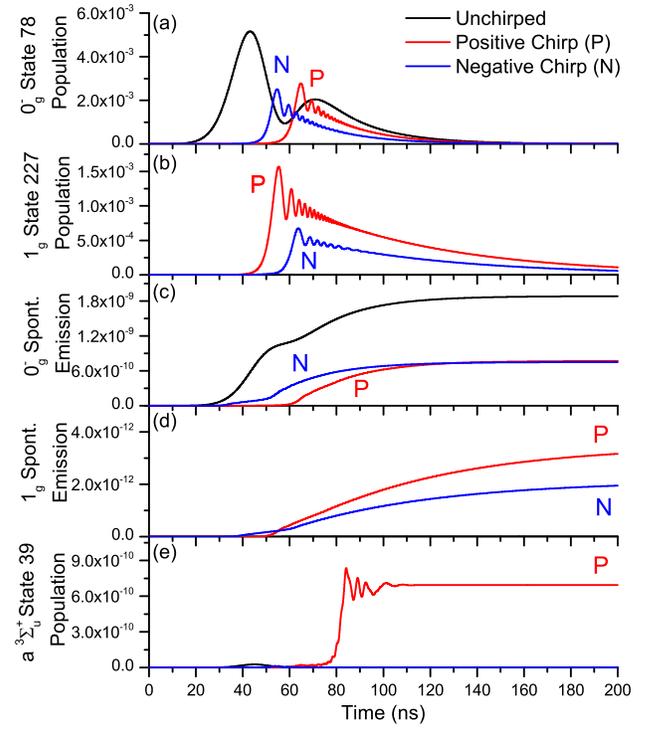} 
    \caption{(Color online) Populations of various molecular states during the unchirped, positively chirped, and negatively chirped pulses for I = 89.29 W/cm$^{2}$: (a) 0$_{g}^{-}$ (v'=78); (b) 1$_{g}$ (v'=227); (c) a $^{3}$$\Sigma$$_{u}^{+}$ bound levels populated by spontaneous emission (SE) from 0$_{g}^{-}$; (d) a $^{3}$$\Sigma$$_{u}^{+}$ bound levels populated by SE from 1$_{g}$; (e) a $^{3}$$\Sigma$$_{u}^{+}$ (v''=39) population resulting from stimulated emission from 0$_{g}^{-}$ (v'=78). Note that in (e), only the positive chirp has a significant contribution.}
\end{figure}

In Fig. 2 we plot these simulated Rb$_{2}$ formation rates vs. peak intensity for various chirped pulses. The values shown correspond to the total number of molecules in a $^{3}$$\Sigma$$_{u}^{+}$ (v''= 0-39) at t=200 ns after the beginning of the chirp. More than 93$\%$ of those molecules reside in v''=37-39 and thus lie within the REMPI bandwidth. The highest level, v''=40, is excluded from this sum because it is bound by only 39 MHz and therefore easily photodissociated by the chirped light. Also, its large outer turning point inhibits detection at our REMPI wavelength. The duration of the chirp is only 100 ns, but we allow for spontaneous emission from the excited states to run its course. Comparing to the experimental results in Fig. 1(b), we see good overall agreement, especially for the dependence on chirp: the rate for positive chirps exceeds that for negative chirps, but is less than that for unchirped pulses.

Comparing chirped and unchirped results is problematic because the pulses have different bandwidths. Ultrafast pulse shaping in the frequency domain \cite{Weiner00} leaves the bandwidth fixed while stretching the pulse. In contrast, our addition of chirp in the time domain maintains the pulse width at 40 ns FWHM, but increases the bandwidth from the transform limit of 11 MHz FWHM to 524 MHz FWHM. In the simulations, we vary the center detuning of the unchirped pulse, with the intensity fixed at 89.29 W/cm$^{2}$, and find a 22 MHz FWHM in the formation rate. Doing the same in the experiment, we find a larger bandwidth of 79 MHz. The limits of the shaded region of Fig. 2 indicate scalings of the unchirped results by the ratio of each of these bandwidths to the chirped bandwidth. This scaling allows a comparison of unchirped and chirped results at the same intensity per unit bandwidth. From this point of view, both the negative and positive chirps are more efficient than unchirped pulses. For completeness, we have also performed the simulations for a much shorter 0.84 ns FWHM unchirped pulse, increasing the peak intensity to keep the pulse energy fixed. The transform limit of this pulse is 524 MHz, the same as for the 40 ns chirped pulses, but its molecular formation rate (dotted curve in Fig. 2) is lower. This is again consistent with higher efficiency for chirped PA at a fixed intensity per unit bandwidth.

The main conclusion to be drawn from Figs. 1(b) and 2 is that the positive chirp gives a higher production rate than the negative chirp. By examining the time dependence of the various state populations, we have identified the mechanism responsible for this difference. In Fig. 3, we show the relative energies of the excited molecular levels involved in the chirp. These are constant in time. We also show the a $^{3}$$\Sigma$$_{u}^{+}$ zero-energy continuum and v''=39 level, with the photon energy added. For clarity, the v''=40 level, bound by only 39 MHz, is not shown. The time dependences of these energies reflect the frequency variations of the chirps: positive in (a) and negative in (b). At a curve crossing, the chirped light is resonant with the transition between the corresponding states. For example, at the peak of the positively-chirped pulse (point B in Fig. 3(a)), the chirp is resonant with the PA transition from the zero-energy continuum to 0$_{g}^{-}$ (v'=78).

In Fig. 4, we plot the time-dependent populations of various excited and a $^{3}$$\Sigma$$_{u}^{+}$ states. Figs. 4(a) and 4(b) show the populations of the two dominant excited states, 0$_{g}^{-}$ (v'=78) and 1$_{g}$ (v'=227), respectively. As expected, the time ordering of population transfer to these states is reversed for the positive and negative chirps. The unchirped pulse excites only to 0$_{g}^{-}$ (v'=78) since it is never resonant with 1$_{g}$ (v'=227). The excited-state populations eventually decay due to spontaneous emission. As shown in Figs. 4(c) and 4(d), a small fraction of these decays populates a $^{3}$$\Sigma$$_{u}^{+}$ high-v'' levels, with 0$_{g}^{-}$ dominating due to better Franck-Condon overlap. Interestingly, as shown in Fig. 4(e), there is another contribution to the v''=39 population, but only for the positive chirp. Referring back to Fig. 3(a), we see that towards the end of the positive chirp (point C), resonance between 0$_{g}^{-}$ (v'=78) and $^{3}$$\Sigma$$_{u}^{+}$ (v''=39) is approached. This results in 0$_{g}^{-}$ (v'=78) population being stimulated \textit{down} to $^{3}$$\Sigma$$_{u}^{+}$ (v''=39). In contrast, for the negative chirp (Fig. 3(b)), this resonance is approached near the beginning of the chirp, when there is no excited population to be stimulated down. The time ordering of these resonances is crucial to the population transfer and breaks the symmetry between positive and negative chirps. If we omit this coherent contribution to the simulated molecular formation rate for the positive chirp, we obtain the dashed curve in Fig. 2, demonstrating that this contribution is responsible for the majority of the difference between positive and negative chirps. The remaining difference is due to the shape variation between positive and negative chirps (Fig. 3). We have verified in the simulations that symmetric linear positive and negative chirps give the same result when this coherent contribution is omitted. We note that this coherent contribution would be even larger if spontaneous emission did not deplete the excited-state population before the stimulated emission occurs.

\section{Conclusion}
In summary, we have investigated the formation of ultracold molecules using frequency-chirped light on the nanosecond timescale. We see a significant enhancement in the formation rate for the positive chirp relative to the negative chirp in both the experimental data and the quantum simulations. The temporal evolutions of the various state populations reveal the mechanism responsible for this difference: photoassociation followed by stimulated emission into a high-vibrational level of the a $^{3}$$\Sigma$$_{u}^{+}$ state. Although we observed a similar trend ($\beta_{pos} > \beta_{neg}$) in the rate constant $\beta$ for trap-loss collisions induced by chirped light with similar parameters \cite{Wright07, Pechkis11, Carini12}, the mechanism here is quite different. The collisional work utilized smaller detunings and thus longer-range excitation, so the time scale of the chirp and the atomic motion were better matched. In the present work, the excited state vibrational period is $\sim$1.7 ns, much shorter than the chirped pulses. We expect that going to faster timescales and higher intensities, together with controlling the details of the chirped pulses \cite{Rogers07}, will allow further optimization of the molecular formation.

\begin{acknowledgments}
	The work at the University of Connecticut is supported by the Chemical Sciences, Geosciences, and Biosciences Division, Office of Basic Energy Sciences, U.S. Department of Energy. 
\end{acknowledgments}


\end{document}